\documentstyle[preprint,aps,epsfig]{revtex}
\begin{document}

\preprint{\vbox{\baselineskip16pt
\hbox{ASITP-2000-002}
\hbox{SNUTP 00-005}
}}
\title{$1/m_Q$ and $1/N_c$ Expansions for Excited Heavy Baryons with Light 
Quarks in the Spin-Flavor Symmetric Representation}

\author{
  Jong-Phil Lee$^a$\footnote{Email address:~jplee@phya.snu.ac.kr},
  Chun Liu$^{b,a}$\footnote{Email address:~liuc@itp.ac.cn}, and
  H.~S. Song$^a$\footnote{Email address:~hssong@physs.snu.ac.kr}}
\address{
  $^a$Department of Physics and Center for Theoretical Physics,\\
  Seoul National University, Seoul 151-742, Korea}
\address{
  $^b$Institute of Theoretical Physics, Chinese Academy of Sciences,\\ 
  P.O. Box 2735, Beijing 100080, China}
\maketitle

\begin{abstract}
The mass spectrum of the $L=1$ orbitally excited heavy baryons with light 
quarks in the spin-flavor symmetric representation is studied by the $1/N_c$ 
expansion method in the framework of the heavy quark effective theory.  The 
mixing effect from the baryons in the mixed representation is considered.
The general pattern of the spectrum is predicted which will be verified by
the experiments in the near future.  The $1/m_Q$ and SU(3) corrections are
also considered.  Mass relations for the baryons $\Lambda_{c1}^{(*)}$, 
$\Sigma_{c1}^{(*)}$, $\Xi_{c1}^{(')(*)}$, and $\Omega_{c1}^{(*)}$ are
derived.
\end{abstract}


A lot of data for orbitally excited heavy baryons have been accumulating 
experimentally \cite{1,2}.  Understanding them will extend our ability in the 
application of QCD.  The heavy quark effective theory (HQET) \cite{3}
provides a systematic way to investigate hadrons containing a single heavy
quark.  To obtain detailed prediction, however, some nonperturbative QCD
methods have to be used.  In this paper, $1/N_c$ expansion \cite{4} is
applied in the analysis.  Within this framework, the masses of $L=1$
orbitally excited heavy baryons with light quarks in both the spin-flavor
symmetric and mixed representation have been analyzed \cite{5}.  By the HQET
sum rule, masses of lowest state of the excited baryons have also been
calculated \cite{6,6a}.  They were studied in other approaches, too, for
example in quark models \cite{7}, in the chiral Lagrangian formalism \cite{8}
and in the Skyrme model or the large $N_c$ \footnote{We make distinction
between $1/N_c$ expansion and the large $N_c$ limit.  Because the former is
essentially based on the light quark spin-flavor symmetry in the baryon
sector, the leading order result of it is not that of $N_c\to\infty$.  See
Manohar, in Ref.\cite{4}.} HQET \cite{9}.  In constituent quark models
\cite{7}, the classification of the baryons according to the light quark
spin-flavor symmetry is taken to be physical.  In the treatment of the
baryons with light quarks in the spin-flavor symmetric representation in
Ref.\cite{5}, it was erroneous to take only one light quark being excited.
In fact, it is the heavy quark that is orbitally excited.  Note that the
orbital excitation of the heavy quark is not suppressed by the mass of the
heavy quark.  Or relatively speaking, it is the light quark pair as a whole
in which the two light quarks have zero relative orbital angular momentum,
that is $L=1$ excited \cite{6,6a,7,8}.  This paper reconsiders the excited
heavy baryons with light quarks in the spin-flavor symmetric representation
in the approach of $1/N_c$ expansion within the framework of the HQET.  The
results are very simple.  Furthermore, the mixing between the two kinds of
representations will also be discussed by $1/N_c$ expansion, which is argued
being small.  Therefore our results are physical and predictive.

In the HQET, many features of heavy hadrons have been analyzed.  In the heavy
quark limit, the heavy quark spin decouples from the strong interaction.  The 
mass of a heavy hadron $H$ is expanded as
\begin{equation}
M_H=m_Q+\bar{\Lambda}_H-\frac{\lambda_1^H}{2m_Q} +\frac{\lambda_2^H}{2m_Q}
    +O(\frac{1}{m_Q^2})\,,
\label{1}
\end{equation}
where $m_Q$ is the heavy quark mass, the parameter $\bar{\Lambda}_H$ is 
independent of the heavy quark spin and flavor, and describes mainly the 
contribution of the light degrees of freedom in the baryon.  $\lambda_1^H$
and $\lambda_2^H$ are the kinetic and chromomagnetic matrix elements, 
respectively,  
\begin{eqnarray}
\lambda^H_1&=&\langle H(v)|\bar{h}_v(iD)^2h_v|H(v)\rangle\,, \nonumber\\
\lambda^H_2&=&-\langle H(v)|\bar{h}_v\frac{g_s}{2}G_{\mu\nu}
              \sigma^{\mu\nu}h_v|H(v)\rangle\,,
\label{2}
\end{eqnarray}
with $h_v$ denoting the heavy quark field with velocity $v$.  The quantities 
$\bar{\Lambda}_H$,  $\lambda_1^H$ and $\lambda_2^H$ should be calculated by 
nonperturbative HQET.  

At this stage, the $1/N_c$ expansion is applied in the analysis.  It is one
of the most important and model-independent methods of nonperturbative QCD.  
Nonperturbative properties of mesons can be observed from the analysis of the 
planar diagrams, and baryons from the Hartree-Fock picture.  For the ground 
state baryons, it has been found that there is a contracted SU(2$N_f$) light 
quark spin-flavor symmetry in the large $N_c$ limit \cite{10,11,12,13}.  This 
makes a $1/N_c$ expansion based on the spin-flavor structure possible for the 
baryons.  Many quantitative predictions and further extensions of the above 
result have been made \cite{14,15,16,17,18,19}.  

Before we go on, two remarks should be made.  First, the above mentioned 
$1/N_c$ expansion applies to the s- or p-wave states of low spin in the
baryon multiplet.  The states with spin of order $N_c/2$ are considerably
modified by spin-spin and spin-orbit interactions \cite{11}.  Second, It is
actually $N_c-1$, which is $2$ in real World, that will be taken as a large
number, because heavy quark is distinguished.  This is an improvement
compared to the $1/N_c$ for the excited heavy baryons with light quarks in
the mixed representation.  In that case, the expansion parameter is $N_c-2$
\cite{5}.

The quantum numbers which describe the hadrons are angular momentum $J$ and
isospin $I$.  For the heavy hadrons, the total angular momentum of the light 
degrees of freedom $J^l$ becomes a good quantum number in the HQET.  In the 
light quark spin-flavor symmetric representation, the light degrees of
freedom in $H$ look like a collection of $N_c-1$ light quarks without orbital
angular momentum excitation.  This picture for the light quarks is
essentially the same as that of the ground state heavy baryons.  The
spin-flavor decomposition rule is $I=S^l$ for the non-strange baryons, where
$S^l$ is the total spin of the light quark system.  Note that the light quark
system as a whole has $L=1$ orbital angular momentum.  In other words, the
heavy quark now is $L=1$ excited in this case.  In real World $N_c$ is fixed
to be 3, so there are only two light quarks in the heavy baryon.  The
spin-flavor structure of them is quite simple, $(I,S^l)=(0,0)$ and $(1,1)$.
All possible states of excited heavy baryons are listed in Table I.  In the
table, except the third state, the other six states form three pairs.  Each
pair is a doublet under the heavy quark spin symmetry.  We adopt the
Hartree-Fock picture to study $\bar{\Lambda}_H$ where in the baryon $H$, the
light quarks are in the spin-flavor symmetric representation.  One of the
essential points of the $1/N_c$ expansion is the $N_c$ counting rules of the
relevant Feynman diagrams. 
 
In the Hartree--Fock picture of the baryons, the $N_c$ counting rules require 
us to include many-body interactions in the analysis, instead of including 
only one- or two-body interactions.  However, a large part of these
interactions are spin-flavor irrelevant.  Namely this part contributes in the
order $N_c\Lambda_{\rm QCD}$ universally to all the baryons with different 
spin-flavor structure in Table I.  This makes us arrive in an $1/N_c$ 
expansion based on the light quark spin-flavor structure of the baryons.  The 
mass splittings among the baryons in the same light quark spin-flavor 
representation can be obtained.  For the purely light quark contribution to 
$\bar{\Lambda}_H$, the $1/N_c$ analysis goes the same as that to the ground 
state heavy baryons \cite{11}.  There is a light quark spin-flavor symmetry
at the leading order of the $1/N_c$ expansion.  $\bar{\Lambda}_H$ is
trivially $\sim N_c\Lambda_{\rm QCD}$ at this order.  The mass splitting due
to the light quark spin-flavor symmetry violation started from ${S^l}^2/N_c$ 
\cite{11}.  However, different from the ground state baryons, formally the
orbital angular momentum of the heavy quark has more dominant contribution to
$\bar{\Lambda}_H$ than $O(1/N_c)$.  This is because of the
orbital-light-quark-spin interactions.  After summing up all the relevant
many-body interactions, this order $O(1)$ contribution is 
$\displaystyle\vec{L}\cdot\vec{S^l}f(\frac{{S^l}^2}{N_c^2})$, where $f$ is a 
general function which can be Taylor expanded.  The mass $\bar{\Lambda}_H$
can be written simply as
\begin{equation}
\bar{\Lambda}^0_H=N_c\tilde{c_0}+\tilde{c_1}\vec{L}\cdot\vec{S^l}
             +\tilde{c_2}\frac{{S^l}^2}{N_c}+O\left(\frac{1}{N_c^2}\right)\,,
\label{3}
\end{equation}
where coefficients $\tilde{c_i}\sim\Lambda_{\rm QCD}$ ($i=0,1,2$).  There 
should be also term proportional to $L^2$ in the above equation, which gives 
constant contribution to $\bar{\Lambda}^0_H$ for a given light quark 
representation, and therefore has been absorbed into the leading term.  The 
term $\vec{L}\cdot\vec{S^l}$ can be rewritten as ${J^l}^2-{S^l}^2$ with 
$\vec{J^l}$ being defined as $\vec{J^l}=\vec{L}+\vec{S^l}$.  Therefore 
\begin{equation} 
\bar{\Lambda}^0_H=N_cc_0+c_1({J^l}^2-{S^l}^2)
                +c_2\frac{{S^l}^2}{N_c}+O\left(\frac{1}{N_c^2}\right)\,,
\label{4}
\end{equation}
where coefficients $c_i\sim\Lambda_{\rm QCD}$ need to be determined from 
experiments.  

The numerical results are also given on the right-handed side of Table I.  
Because the mass formula of Eq. (\ref{4}) is rather simple, some features of
the spectrum can still be discussed.  The parameters $c_0$ and $c_2$ are
naturally expected to be positive.  However $c_1$ can have both signs.  If
$c_1>0$, we see that the singlet state ($J$,$I$) $=$ ($\frac{1}{2}$,1) could
be the lowest state.  By requiring the first doublet to be the lowest, we
must have $c_2 > 2N_cc_1$.  The resulting spectrum will be 
\begin{equation} 
M(\frac{1}{2}(\frac{3}{2}), 0, 1, 0) < M(\frac{1}{2}, 1, 0, 1) < 
M(\frac{1}{2}(\frac{3}{2}), 1, 1, 1) < M(\frac{3}{2}(\frac{5}{2}), 1, 2, 1) 
\label{a}
\end{equation}
with the quantum numbers denoting $J$, $I$, $J^l$ and $S^l$, respectively.  
On the other hand, if $c_1 < 0$, the first doublet is the lowest states only 
if $c_2 > -N_cc_1$.  In this case, the singlet is the heaviest, and the 
spectrum is
\begin{equation} 
M(\frac{1}{2}(\frac{3}{2}), 0, 1, 0) < M(\frac{3}{2}(\frac{5}{2}), 1, 2, 1) 
< M(\frac{1}{2}(\frac{3}{2}), 1, 1, 1) < M(\frac{1}{2}, 1, 0, 1)\,. 
\label{b}
\end{equation}

None of the above discussed spectrum pattern is consistent with the quark 
model prediction \cite{7}.  It should be noted that our analysis neglected
the $1/N_c^3$ correction (compared to the leading order) which is expected to
be not significant.

The conditions for $c_2$ are not satisfactory, although they are not
unreasonable considering that in real World $N_c$ is not large.  In fact,
this unsatisfactory point can be avoided if the mixing effect from the
baryons in the mixed representation is considered.

It is necessary to consider the mixing between the baryons with light quarks
in the spin-flavor symmetric and mixed representations.  When they have same
quantum numbers of ($J$, $I$, $J^l$), there is no physical way to distinguish
them.  This consideration will give the physical spectrum.  Because of the
light quark spin-flavor symmetry at the leading order of $1/N_c$ expansion,
the baryons with same ($J$, $I$, $J^l$) quantum numbers but in different
representations do not mix.  The mixing occurs at the sub-leading order.  The
classification of baryons by the spin-flavor symmetry  is therefore physical
at the leading order \cite{21}.  For the physical spectrum, the mixing
results in a deviation from $\bar{\Lambda}^0_H$.  By denoting the mixing mass
as $\tilde{m}$ which is of $O(1)$, the mass matrix for the baryons with same
($J, I, J^l$) is written as
\begin{equation}
\label{c}
\left(\begin{array}{cc}
\bar{\Lambda}^0_H & \tilde{m}\\
\tilde{m}         & \bar{\Lambda}^0_{H'}\\
\end{array}
\right)
\,,
\end{equation}
where $H'$ is the corresponding baryon in the mixed representation.  
$\bar{\Lambda}^0_{H'}$ was given in Ref. \cite{5}.  The mass difference
$\bar{\Lambda}^0_H-\bar{\Lambda}^0_{H'}$ is $O(1)$.  Taking
$\tilde{m}<\bar{\Lambda}^0_{H'}-\bar{\Lambda}^0_H$ for illustration, the
physical mass are corrected to be
\begin{equation}
\begin{array}{lll}
\bar{\Lambda}_H   &\simeq&\displaystyle\bar{\Lambda}^0_H-\frac{\tilde{m}^2}
{\bar{\Lambda}^0_{H'}-\bar{\Lambda}^0_H}~,\\[3mm]
\bar{\Lambda}_{H'}&\simeq&\displaystyle\bar{\Lambda}^0_{H'}
+\frac{\tilde{m}^2}{\bar{\Lambda}^0_{H'}-\bar{\Lambda}^0_H}~.
\label{c1}
\end{array}
\end{equation}
The mixing effect
$\displaystyle\frac{\tilde{m}^2}{\bar{\Lambda}^0_{H'}-\bar{\Lambda}^0_H}$ is
positive.  It reduces the predictive power of Eq. (\ref{4}) for the mass
spectrum.  The $1/N_c$ expansion of $\tilde{m}$ is parameterized as
\begin{equation}
\label{c2}
\tilde{m}=\tilde{m}_0+O(1/N_c)\,,
\end{equation}
where $\tilde{m}_0$ is universal due to the light quark spin-flavor symmetry.
To the order of $O(1)$, the spectrum is given as follows explicitly.
\begin{equation}
\begin{array}{lll}
\displaystyle\bar{\Lambda}_{(\frac{1}{2}(\frac{3}{2}), 0, 1)}&=&\displaystyle
N_cc_0+2c_1-\frac{\tilde{m}_0^2}
{k-c_{LS}-\frac{1}{6}\bar{c_1}-\frac{1}{4}\bar{c_2}-2c_1}~,\\[3mm]
\displaystyle\bar{\Lambda}_{(\frac{1}{2}, 1, 0)}      &=&N_cc_0-2c_1~,\\[3mm]
\displaystyle\bar{\Lambda}_{(\frac{1}{2}(\frac{3}{2}), 1, 1)}&=&\displaystyle
N_cc_0-\frac{\tilde{m}_0^2}{k}~,\\[3mm]
\displaystyle\bar{\Lambda}_{(\frac{3}{2}(\frac{5}{2}), 1, 2)}&=&N_cc_0+4c_1~,
\label{c3}
\end{array}
\end{equation}
where $k$ is an $O(1)$ constant that remains after the
$\bar{\Lambda}^0_{H'}$ and $\bar{\Lambda}^0_H$ cancellation,
$\bar{\Lambda}^0_{H'}$ is parameterized by $c_{LS}$, $\bar{c_1}$ and
$\bar{c_2}$ which are around $\Lambda_{\rm QCD}$, and can be found in the
Table II of Ref. \cite{5} (where $\bar{c_1}$ and $\bar{c_2}$ were denoted as
$c_1$ and $c_2$, respectively).  Note that the masses of the states
$\displaystyle (\frac{1}{2}, 1, 0)$ and
$\displaystyle (\frac{3}{2}(\frac{5}{2}), 1, 2)$ are not affected by the
mixing, because there are no physical states with the same good quantum
numbers in the mixed representation.  From the above spectrum, we see that
$c_1>0$.  The states $\displaystyle (\frac{3}{2}(\frac{5}{2}), 1, 2)$ is
always the highest states.  They are heavier than the other states at least
by $4c_1$ through requiring the states
$\displaystyle (\frac{1}{2}(\frac{3}{2}), 0, 1)$ to be the lowest.  If
$\displaystyle 2c_1>\frac{\tilde{m}_0^2}{k}$, the requirement implies
\begin{equation}
\label{c4}
\frac{\tilde{m}_0^2}
{k-c_{LS}-\frac{1}{6}\bar{c_1}-\frac{1}{4}\bar{c_2}-2c_1}>4c_1~.
\end{equation}
In this case, the spectrum pattern is
\begin{equation} 
M(\frac{1}{2}(\frac{3}{2}), 0, 1)<M(\frac{1}{2}, 1, 0)< 
M(\frac{1}{2}(\frac{3}{2}), 1, 1)<M(\frac{3}{2}(\frac{5}{2}), 1, 2)\,. 
\label{c5}
\end{equation}
On the other hand, if $\displaystyle 2c_1<\frac{\tilde{m}_0^2}{k}$, the
requirement is
\begin{equation}
\label{c6}
\tilde{m}_0^2\left(\frac{1}
{k-c_{LS}-\frac{1}{6}\bar{c_1}-\frac{1}{4}\bar{c_2}-2c_1}-\frac{1}{k}\right)
>2c_1~,
\end{equation}
which gives the spectrum
\begin{equation} 
M(\frac{1}{2}(\frac{3}{2}), 0, 1)<M(\frac{1}{2}(\frac{3}{2}), 1, 1)<
M(\frac{1}{2}, 1, 0)<M(\frac{3}{2}(\frac{5}{2}), 1, 2)\,.
\label{c7}
\end{equation}

Experimentally, the excited charmed baryons $\Lambda_{c1}(\frac{1}{2})$ and 
$\Lambda_{c1}(\frac{3}{2})$ have been found which correspond to the 
$\displaystyle (\frac{1}{2}(\frac{3}{2}), 0, 1)$ states.  More data are
needed to fix the unknown parameters $c_i$'s, $\bar{c_i}$'s, $k$ and
$c_{LS}$.  In the near future, experiments will check the above predicted
spectrum.  Hopefully one of the above mass patterns will be picked out.  It
will be a check for the validity of our method, if the parameters are in the
reasonable range ($\Lambda_{\rm QCD}$) and meanwhile satisfy the relations
given above.

For a complete analysis of the heavy hadron masses, $1/m_Q$ corrections have 
to be considered.  The general expression of the corrections have been given 
in Eqs. (\ref{1}) and (\ref{2}).  The quantities $\lambda^H_1$ and 
$\lambda^H_2$ can be analyzed by the $1/N_c$ expansion in the similar way as 
$\bar{\Lambda}_H$.  In the leading order of $1/N_c$, $\lambda^H_1$ is 
independent of the light quark structure and scales as unity.  Therefore we 
have the following expansion ,
\begin{eqnarray}
\lambda^H_1&=&\displaystyle\tilde{c_0}'+\tilde{c_1}'\vec{L}\cdot 
              \frac{\vec{S^l}}{N_c}
              +O\left(\frac{1}{N_c^2}\right)\nonumber\\[3mm] 
           &=&\displaystyle c_0'+c_1'\frac{{J^l}^2-{S^l}^2}{N_c}
              +O\left(\frac{1}{N_c^2}\right)\,.
\label{7}
\end{eqnarray}
The mixing effect also affects $\lambda^H_1$.  Its $1/N_c$ expansion is that
the non-vanishing contribution begins at $O(1/N_c)$.  And at this order, the
contribution is constant which can be absorbed into $c_0'$.  The parameters
$c_0'/2m_Q$ and $c_1'/2m_Q$ can be absorbed into $c_0$ and $c_1$ in Eq.
(\ref{4}), respectively.  The inclusion of $\lambda^H_1$ corrects the masses
of the baryons at the order of $1/m_Q$ which is expected to be not
significant.  It does not change the mass pattern given above to the order of
$O(1/m_cN_c)$.  

The degeneracy in the spectrum due to the heavy quark spin symmetry is lifted 
by $\lambda^H_2$.  According to the definition in Eq. (\ref{2}),
$\lambda^H_2$ is heavy baryon spin dependent.  It is convenient to extract
this dependence explicitly,
\begin{equation}
\lambda^H_2=d_H\lambda_2
\label{8}
\end{equation}
where $d_H=2j^l$ for $H$ with $J=j^l+\frac{1}{2}$, and $d_H=-2j^l-2$ for $H$ 
with $J=j^l-\frac{1}{2}$.  The new defined heavy quark hadronic matrix
element $\lambda_2$ is heavy baryon spin independent.  It is also independent
of the light quark structure and scales as unity in the leading order $1/N_c$ 
expansion.  Like $\lambda^H_1$, the $1/N_c$ expansion for $\lambda^H_2$ is
\begin{equation}
\lambda^H_2=d_H\left[c_0''+c_1''\frac{{J^l}^2-{S^l}^2}{N_c}
              +O\left(\frac{1}{N_c^2}\right)\right]\,.\\
\label{9}
\end{equation}
The mixing effect for $\lambda^H_2$ is that the leading nonzero contribution
is $O(1/N_c)$ which is constant and therefore can be absorbed into $c_0"$.  
The parameters $c_i''$ should be determined by the experimental data.  If we
work to the accuracy of $\Lambda_{\rm QCD}/(m_QN_c)\sim 10\%$, $c_0''$ can be 
fixed from the mass splitting of $\Lambda_{c1}(\frac{3}{2})$ and
$\Lambda_{c1}(\frac{1}{2})$, 
\begin{equation}
c_0''=\frac{m_c}{3}\left[M_{\Lambda_{c1}(\frac{3}{2})}
      -M_{\Lambda_{c1}(\frac{1}{2})}\right]\simeq (128~~{\rm MeV})^2\,,
\label{10}
\end{equation}
by taking $m_c\simeq 1.5$ GeV.  Note that $c_0''$ is positive.  The mass 
splittings of the other degenerate states listed in Table I are predicted to 
be
\begin{eqnarray}
M(\frac{5}{2}, 1, 2, 1)-M(\frac{3}{2}, 1, 2, 1)&=&
\displaystyle\frac{5c_0''}{m_c}\simeq 55~~~{\rm MeV}\,,\nonumber\\[3mm]
M(\frac{3}{2}, 1, 1, 1)-M(\frac{1}{2}, 1, 1, 1)&=&
\displaystyle\frac{3c_0''}{m_c}\simeq 33~~~{\rm MeV}\,,
\label{11}
\end{eqnarray}
to the accuracy of $c_0''^2/(m_cN_c)$ which is about $5$ MeV.  These 
predictions can be checked with the experiments in the near future.

Finally, let us consider the case of the excited heavy baryons with light 
quarks including the strange quark.  Very recently, there are experimental 
evidence of the charmed-strange analogs of $\Lambda_{c1}(\frac{3}{2})$, 
$\Xi_{c1}(\frac{3}{2})$ particles \cite{2}.  The above framework can be
easily extended to include the charmed-strange baryons by taking strangeness
as perturbation to the light quark flavor symmetry.  The relevant baryon mass
is then expressed as 
\begin{equation}
M_H=m_Q+N_cc_0+c_1({J^l}^2-{S^l}^2)+{\rm mixing}+c_3(-s)
    +O\left(\frac{1}{N_c}\right)\,,\\
\label{12}
\end{equation}
where $s$ is the heavy baryon strangeness number which can be $0$, $-1$, or
$-2$.  The parameter $c_3$ stands for the leading order of SU(3) correction to 
the $\bar{\Lambda}_H$ given in Eq. (\ref{4}).  It is fixed by the mass 
difference of $\Xi_{c1}(\frac{3}{2})$ and $\Lambda_{c1}(\frac{3}{2})$,
\begin{equation}
c_3\simeq 190~~{\rm MeV}\,.
\label{13}
\end{equation}
The mass of $\Xi_{c1}(\frac{1}{2})$ is then predicted to be 190 MeV higher 
than $\Lambda_{c1}(\frac{1}{2})$,
\begin{eqnarray}
M_{\Xi_{c1}(\frac{1}{2})}&=&M_{\Lambda_{c1}(\frac{1}{2})}
           +M_{\Xi_{c1}(\frac{3}{2})}-M_{\Lambda_{c1}(\frac{3}{2})}\nonumber\\
                         &\simeq& 2784~~{\rm MeV}\,.
\label{14}
\end{eqnarray}
Note that this prediction is only subject to a small uncertainty which is 
about $c_3^2/(m_cN_c)\sim 10$ MeV.  The future experiments may find that 
particles $\Sigma_{c1}^{(*)}$, which are the lowest charmed $L=1$ states with 
isospin $1$, are just the state pair $[\frac{1}{2}(\frac{3}{2}), 1, 1, 1]$ in 
Table I.  Their strange analogs $\Xi_{c1}'^{(*)}$ and $\Omega_{c1}^{(*)}$ will 
then be predicted from the similar relation rather precisely,
\begin{eqnarray}
M_{\Xi_{c1}'(\frac{3}{2})}-M_{\Sigma_{c1}(\frac{3}{2})}&=&
M_{\Xi_{c1}'(\frac{1}{2})}-M_{\Sigma_{c1}(\frac{1}{2})}
+O\left(\frac{\Lambda_{\rm QCD}}{m_cN_c}\right) \,,\nonumber\\
M_{\Omega_{c1}(\frac{3}{2})}-M_{\Sigma_{c1}(\frac{3}{2})}&=&
M_{\Omega_{c1}(\frac{1}{2})}-M_{\Sigma_{c1}(\frac{1}{2})}
+O\left(\frac{\Lambda_{\rm QCD}}{m_cN_c}\right) \,.
\label{15}
\end{eqnarray}
To the accuracy of $s^2/N_c\sim 30\%$,
\begin{eqnarray}
M_{\Omega_{c1}^{(*)}}-M_{\Xi_{c1}'^{(*)}}
&=&M_{\Xi_{c1}'^{(*)}}-M_{\Sigma_{c1}^{(*)}}
+O\left(\frac{s^2}{N_c}\right)\nonumber\\ 
&\simeq&(190\pm70)~~{\rm MeV}\,.
\label{16}
\end{eqnarray}

In summary, we have applied the $1/N_c$ expansion method to study the mass 
spectrum of the $L=1$ orbitally excited heavy baryons with light quarks being 
in the spin-flavor symmetric representation within the framework of the HQET.  
The analysis is very simple compared to that for the heavy baryons with light 
quarks in the mixed representation in Ref. \cite{5}.  The simplicity is an 
unique feature in this case.  It can be seen from the following point of
view, namely the light quark system is in the ground state and it is the
heavy quark that is orbitally excited.  However the mixing effect due to the
baryon states in the mixed representation corrects the spectrum pattern in
the subleading order of $1/N_c$ expansion.  The effect is important to get
the realistic spectrum at this order.  The general pattern of the baryon
spectrum has been given, which will be verified by the experiments in the
near future.  The $1/m_Q$ and SU(3) corrections have also been considered.
Certain mass relations for the baryons $\Lambda_{c1}^{(*)}$,
$\Sigma_{c1}^{(*)}$, $\Xi_{c1}^{(')(*)}$, and $\Omega_{c1}^{(*)}$ have been
derived.  The same analysis can be applied to the bottom baryons.  

\begin{center}
{\large\bf Acknowledgments}\\[10mm]
\end{center}\par

We would like to thank Chao-Hsi Chang and Chao-shang Huang for helpful
discussions.  This work was supported in part by the BK21 Program of the
Korean  Ministry of Education.


\newpage

\begin{table}
Table I.  Excited heavy baryon states of the symmetric representation of 
$N_c-1$ light quarks.  The masses are that without considering the mixing.\\
\vskip 5.3mm
\begin{center}
\begin{tabular}{ccc}
$(J,I)$   & $(J^l,S^l)$ & $\bar{\Lambda}_H$\\\hline
$(1/2,0)$ & $(1,0)$     & $N_cc_0+2c_1$\\
$(3/2,0)$ & $(1,0)$     & $N_cc_0+2c_1$\\
$(1/2,1)$ & $(0,1)$     & $N_cc_0-2c_1+\frac{2c_2}{N_c}$\\
$(1/2,1)$ & $(1,1)$     & $N_cc_0+\frac{2c_2}{N_c}$\\
$(3/2,1)$ & $(1,1)$     & $N_cc_0+\frac{2c_2}{N_c}$\\
$(3/2,1)$ & $(2,1)$     & $N_cc_0+4c_1+\frac{2c_2}{N_c}$\\
$(5/2,1)$ & $(2,1)$     & $N_cc_0+4c_1+\frac{2c_2}{N_c}$
\end{tabular}
\end{center}
\label{table}
\end{table}

\end{document}